\documentclass[12pt]{article}
\usepackage{amssymb}
\newcommand{\be}{\begin{eqnarray}}
\newcommand{\ee}{\end{eqnarray}}
\newcommand{\pa}{\partial}
\renewcommand{\d}{{\rm d}}

\title{\bf A noncommutative version of the nonlinear Schr\"odinger equation}

\date{  }

\author{A. Dimakis \\ Department of Mathematics, University of the Aegean \\
        GR-83200 Karlovasi, Samos, Greece \\ dimakis@aegean.gr
        \\[2ex]
        F. M\"uller-Hoissen \\ Max-Planck-Institut f\"ur Str\"omungsforschung \\
        Bunsenstrasse 10, D-37073 G\"ottingen, Germany \\
        fmuelle@gwdg.de }

\begin{document}

\renewcommand{\theequation} {\arabic{section}.\arabic{equation}}

\maketitle

\begin{abstract}
We apply a (Moyal) deformation quantization to a bicomplex associated with 
the classical nonlinear Schr\"odinger equation. This induces a deformation 
of the latter equation to noncommutative space-time while preserving the 
existence of an infinite set of conserved quantities.
\end{abstract}

\section{Introduction}
\setcounter{equation}{0}
Partly motivated by recent results on the appearance of field theories on 
noncommutative space-times in certain limits of string, D-brane and M theory 
\cite{ncg_str}, there has been a revival of (Moyal) deformation quantization 
\cite{dq} and increasing interest in models on noncommutative spaces 
(see \cite{dq_nc}, for example). 
\vskip.1cm

In this work we apply deformation quantization to a classical integrable 
model\footnote{Deformation quantization in the context of integrable models 
has been considered in \cite{def_im,Stra97,nc_im}, for example.}, 
the nonlinear Schr\"odinger equation. More precisely, we apply it to the two 
space-time coordinates (see also \cite{nc_im}) and generalize a bicomplex 
associated with the classical nonlinear partial differential equation to 
the resulting ``quantized space-time". The deformed bicomplex equations 
are then equivalent to a deformed nonlinear Schr\"odinger equation which 
still possesses an infinite set of conservation laws. It is this latter 
characteristic property of soliton equations which distinguishes the resulting 
``quantized equation" and turns it into an interesting model.
\vskip.1cm

Section 2 recalls some bicomplex formalism which has been developed in \cite{DMH00a}. 
In section 3 we treat the nonlinear Schr\"odinger equation in this framework. 
Section 4 deals with the corresponding noncommutative extension and in section 5 
we show that the single soliton solution of the classical equation is also a solution 
of the noncommutative version. Section 6 deals with perturbative properties 
of this equation and section 7 contains some conclusions.

\section{Bicomplexes and associated linear equation}
\setcounter{equation}{0}
A {\em bicomplex} is an ${\mathbb{N}}_0$-graded linear space (over ${\mathbb{R}}$ 
or ${\mathbb{C}}$) $ M = \bigoplus_{r \geq 0} M^r $ together with
two linear maps $\d , \delta \, : \, M^r \rightarrow M^{r+1}$ 
satisfying
\be
   \d^2 = 0 \, , \qquad  \delta^2 = 0 \, , \qquad  
   \d \, \delta + \delta \, \d = 0 \; .   \label{bicomplex_cond}
\ee 
\vskip.2cm

Let us assume that there is a (nonvanishing) $\chi^{(0)} \in M^0$ with 
$\d J^{(0)} =0$ where $J^{(0)} = \delta \chi^{(0)}$. 
Let us define $J^{(1)} = \d \chi^{(0)}$. Then 
$\delta J^{(1)} = - \d \delta \chi^{(0)} = 0$. Hence $J^{(1)}$ is $\delta$-closed. 
If $J^{(1)}$ is $\delta$-exact, then $J^{(1)} = \delta \chi^{(1)}$ 
with some $\chi^{(1)} \in M^0$. Next we define
$J^{(2)} = \d \chi^{(1)}$. Then 
$\delta J^{(2)} = - \d \delta \chi^{(1)} = - \d J^{(1)} = - \d^2 \chi^{(0)} = 0$. 
If the $\delta$-closed element $J^{(2)}$ is $\delta$-exact, then 
$J^{(2)} = \delta \chi^{(2)}$ with some $\chi^{(2)} \in M^0$. 
This can be iterated further and leads to a (possibly infinite) 
chain of elements $J^{(m)}$ of $M^1$ and 
$\chi^{(m)} \in M^0$ satisfying
\be
         J^{(m+1)} = \d \chi^{(m)} = \delta \chi^{(m+1)}   \; .
\ee
More precisely, the above iteration continues from the $m$th to the
$(m+1)$th level as long as $\delta J^{(m)} = 0$ 
implies $J^{(m)} = \delta \chi^{(m)}$ with an element 
$\chi^{(m)} \in M^0$. Of course, there is no obstruction to the 
iteration if $H^1_\delta (M)$ is trivial, 
i.e., when all $\delta$-closed elements of $M^1$ are $\delta$-exact.
In general, the latter condition is too strong, however.
Introducing 
\be
   \chi = \sum_{m \geq 0} \lambda^m \, \chi^{(m)}  
          \label{chi-ps}
\ee
as a (formal) power series in a parameter $\lambda$, the essential  
steps of the above iteration procedure are summarized in
\be
    \delta (\chi - \chi^{(0)}) = \lambda \, \d \, \chi  
               \label{bc-linear}
\ee
which we call the {\em linear equation} associated with the bicomplex.
\vskip.2cm

Given a bicomplex, we may start directly with the linear 
equation (\ref{bc-linear}). Let us assume that it admits a (non-trivial)
solution $\chi$ as a (formal) power series (\ref{chi-ps}) in $\lambda$.
The linear equation then leads to 
\be
   \delta \chi^{(m)} = \d \chi^{(m-1)} \, , \quad m=1, \ldots, \infty \; .
\ee 
As a consequence, the $J^{(m+1)} = \d \chi^{(m)}$ ($m=0, \ldots, \infty$) are 
$\delta$-exact. Even if the cohomology $H^1_\delta (M)$ is {\em not} trivial, 
solvability of the linear equation ensures that the $\delta$-closed $J^{(m)}$
appearing in the iteration are $\delta$-exact.
\vskip.2cm

 For some integrable models the ``generalized conserved currents" $J^{(m)}$ of 
an associated bicomplex are directly related to conserved densities \cite{DMH00a}. 
In other cases, like the nonlinear Schr\"odinger equation treated in the following 
section, the relation is less direct.

\section{Bicomplex formulation of the classical nonlinear \\ Schr\"odinger equation}
\setcounter{equation}{0}
We choose the bicomplex space as $M = M^0 \otimes \Lambda$ where $M^0 = C^\infty({\mathbb{R}^2},{\mathbb{C}}^2)$ denotes the set of smooth 
maps $\phi : {\mathbb{R}}^2 \rightarrow {\mathbb{C}}^2$ and 
$\Lambda = \bigoplus_{r=0}^2 \Lambda^r$ is the exterior algebra of a 
$2$-dimensional complex vector space with basis $\tau,\xi$ 
of $\Lambda^1$ (so that $\tau^2 = \xi^2 = \tau \, \xi + \xi \, \tau = 0$). 
It is then sufficient to define bicomplex maps $\d$ and $\delta$ on $M^0$ 
since by linearity and $\d (\phi_1 \, \tau + \phi_2 \, \xi) 
= (\d \phi_1) \, \tau + (\d \phi_2) \, \xi$ (and correspondingly for $\delta$) 
they extend to the whole of $M$. Let 
\be
    \d \phi &=& (\phi_t - V \, \phi) \, \tau + (\phi_x - U \, \phi) \, \xi \\
    \delta \phi &=& \phi_x \, \tau + {1 \over 2i} \, (I - \sigma_3) \, \phi \, \xi
\ee
where $I$ is the $2 \times 2$ unit matrix, $\sigma_3 = \mbox{diag}(1,-1)$, and
\be
    U = \left( \begin{array}{cc} 0 & -\bar{\psi} \\ \psi & 0  \end{array} \right) 
          \, , \quad
    V = i \, \left( \begin{array}{cc} -|\psi|^2 & \bar{\psi}_x \\
                     \psi_x & |\psi|^2 
              \end{array} \right) 
\ee
with a complex function $\psi$ with complex conjugate $\bar{\psi}$.
Then $\delta^2 =0$ and $\d \delta + \delta \d = 0$ are 
identically satisfied. Furthermore, $\d^2 = 0$ holds iff 
$U_t - V_x + [ U , V ] = 0$ which is equivalent to the nonlinear 
Schr\"odinger equation
\be
    i \, \psi_t = - \psi_{xx} - 2 \, |\psi|^2 \, \psi  \; .
\ee
\vskip.2cm

The equation $\delta J = 0$ for an element 
$J = \phi_1 \, \tau + \phi_2 \, \xi \in M^1$ means 
$\phi_{2,x} = - (i/2) (I-\sigma_3) \phi_1$. 
Let $\Phi \in M^{0}$ such that $\Phi_x = \phi_1$. Then 
$\phi_2 = -(i/2) (I-\sigma_3) \Phi + C$ 
with an element $C(t) \in M^{0}$ and we find 
$ J = \Phi_x \, \tau - [(i/2)(I-\sigma_3) \Phi - C] \, \xi 
  = \delta \Phi + C \, \xi $. 
The cohomology of $\delta$ is not trivial since an element $(c(t),0)^T \xi$ 
is $\delta$-closed, but obviously not $\delta$-exact.\footnote{$\phi^T$ 
denotes the transpose of $\phi$.} 
\vskip.2cm

Choosing $\chi^{(0)}$ such that\footnote{The general solution of 
$\delta \chi^{(0)} =0$ is $(c(t),0)^T$.} 
$\delta \chi^{(0)} =0$, the linear equation associated with the 
bicomplex becomes
\be
    \chi_x = \lambda \, ( \chi_t - V \chi) \, , \qquad
    (I-\sigma_3) \chi = 2 \, i \, \lambda \, (\chi_x - U \chi)  \; .
\ee
In terms of $ \chi = (\alpha , \beta )^T $ this system of equations 
takes the form
\be
   \alpha_x &=& \lambda \, \{ \alpha_t + i \, |\psi|^2 \alpha 
                - i \, \bar{\psi}_x \beta \} \\
   \beta_x &=& \lambda \, \{ \beta_t - i \, \psi_x \, \alpha 
               - i \, |\psi|^2 \beta \} \\
   \alpha_x &=& - \bar{\psi} \, \beta   \label{beta_eq}  \\
   \beta &=& i \, \lambda \, ( \beta_x - \psi \, \alpha ) \; .
\ee
The third equation can be used to eliminate $\beta$ in the other equations (assuming $\psi \neq 0$). We obtain
\be
   \alpha_x &=& \lambda \, \{ \alpha_t + i \, |\psi|^2 \alpha 
                + i \, (\bar{\psi}_x/\bar{\psi}) \, \alpha_x \}  
                          \label{alpha_eq1}   \\
   \alpha_x &=& i \, \lambda \, ( \alpha_{xx} - (\bar{\psi}_x/\bar{\psi}) \alpha_x 
                + |\psi|^2 \alpha )   \; .   \label{alpha_eq2}
\ee
Conversely, if these two equations hold and if $\psi$ satisfies the nonlinear
Schr\"odinger equation, then the linear equation is satisfied (where $\beta$ is
defined via (\ref{beta_eq})). Moreover, the above two equations for $\alpha$ 
are compatible, i.e., the equation obtained by differentiating the second equation 
with respect to $t$ is identically satisfied as a consequence of these equations 
together with the nonlinear Schr\"odinger equation.
\vskip.2cm

Now we choose $\chi^{(0)} = (1,0)^T$. In terms of $\gamma$ determined by
\be 
   \alpha = e^{i \lambda \gamma} \, , \qquad 
   \gamma = \sum_{m \geq 0} \lambda^m \, \gamma^{(m)}  
            \label{alpha_gamma}
\ee
the last two equations read
\be
   \gamma_x &=& |\psi|^2 + i \, \lambda \, [ \gamma_{xx}  
                - (\bar{\psi}_x/\bar{\psi}) \gamma_x ]
                - \lambda^2 (\gamma_x)^2  \label{gamma1}  \\
   \gamma_t &=& i \, [ \gamma_{xx} - 2 \, (\bar{\psi}_x/\bar{\psi}) \, \gamma_x ]
                - \lambda \, (\gamma_x)^2    \; .
                                          \label{gamma2} 
\ee
Differentiation of the second equation with respect to $x$ leads to the 
conservation law
\be
   \gamma_{xt} &=& [ i \, \gamma_{xx} - 2 \, i \, (\bar{\psi}_x/\bar{\psi}) 
                   \, \gamma_x - \lambda \, (\gamma_x)^2 ]_x 
\ee
for $ \gamma_x $.
Inserting the power series expansion for $\gamma$ in (\ref{gamma1}), we obtain
\be
   \gamma^{(0)}_x = |\psi|^2 \, , \qquad 
   \gamma^{(1)}_x = i \, \bar{\psi} \, \psi_x    \label{gamma_rec1}
\ee
and the recursion formula
\be
    \gamma^{(m)}_x = i \, [ \gamma^{(m-1)}_{xx}  
                - (\bar{\psi}_x/\bar{\psi}) \, \gamma^{(m-1)}_x ]
                - \sum_{k=0}^{m-2} \gamma^{(k)}_x \, \gamma^{(m-2-k)}_x
                        \label{gamma_rec2}
\ee
for $m > 1$. In particular, 
$\gamma^{(2)}_x = - |\psi|^4 - \bar{\psi} \, \psi_{xx}$.
These are the well-known conserved quantities of the nonlinear 
Schr\"odinger equation (cf \cite{Fadd+Takh87}, p.38, for example). 
(\ref{gamma_rec1}) and (\ref{gamma_rec2}) solve (\ref{gamma1}). 
(\ref{gamma2}) becomes 
$\gamma^{(0)}_t = i \, [ \gamma^{(0)}_{xx} - 2 \, (\bar{\psi}_x/\bar{\psi}) 
                  \, \gamma^{(0)}_x ]$ 
and 
\be
   \gamma^{(m)}_t = i \, [ \gamma^{(m)}_{xx} - 2 \, (\bar{\psi}_x/\bar{\psi}) 
                    \, \gamma^{(m)}_x ]
                - \sum_{k=0}^{m-1} \gamma^{(k)}_x \, \gamma^{(m-1-k)}_x
\ee
for $m>0$.
Inserting our solution $\gamma^{(m)} = \int^x \gamma^{(m)}_{x'} \, dx'$ with
the $\gamma^{(m)}_x$ as given above, these equations are satisfied as a 
consequence of the nonlinear Schr\"odinger equation (which in turn is a 
consequence of the compatibility of (\ref{gamma1}) and (\ref{gamma2}), 
respectively (\ref{alpha_eq1}) and (\ref{alpha_eq2})).

\section{Nonlinear Schr\"odinger equation in noncommutative space-time}
\setcounter{equation}{0}
The passage from commutative to noncommutative space-time is (in the present 
context) achieved by replacing the ordinary commutative product in the space 
of smooth functions on $\mathbb{R}^2$ with coordinates $t,x$ by the 
noncommutative associative (Moyal) $\ast$-product \cite{dq} which is 
defined by
\be
   f \ast h = m \circ e^{\vartheta P/2} (f \otimes h)   \label{ast}
\ee
where $\vartheta$ is a real or imaginary constant and 
\be
   m (f \otimes h) = f \, h \, , \qquad
   P = \pa_t \otimes \pa_x - \pa_x \otimes \pa_t  \; .
\ee
As a consequence, one finds\footnote{This is easily verified using
$\tau (f \otimes h) = h \otimes f$ which has the properties $P \circ \tau
= - \tau \circ P$ and $m \circ \tau = m$.}
\be
   f \ast h - h \ast f = 2 \, m \circ \sinh (\vartheta P/2) (f \otimes h)
   = \vartheta \, m \circ
   \frac{\sinh(\vartheta P/2)}{\vartheta P/2} \circ P (f \otimes h) \; .
      \label{ast-comm}
\ee
In the following, the product\footnote{This product has been 
previously used in \cite{Stra97}.}
defined by 
\be
   f \diamond h 
 = m \circ \frac{\sinh(\vartheta P/2)}{\vartheta P/2}(f \otimes h)
    \label{diam}
\ee
will be of some help for us.
\vskip.2cm

In noncommutative space-time, the two bicomplex maps associated with the 
nonlinear Schr\"odinger equation should be replaced by
\be
  \d \phi &=& (\phi_t - V \ast \phi) \, \tau + (\phi_x - U \ast \phi) \, \xi  \\
  \delta \phi &=& \phi_x \, \tau  + {1 \over 2i} (I-\sigma_3) \phi \, \xi
\ee
with $2 \times 2$ matrices $U$ and $V$. 
The bicomplex space $M$ is the same as in the previous section, 
however. For the following calculations it is important to note that partial 
derivatives are also derivations with respect to the $\ast$-product. 
The bicomplex conditions imply
\be
   U_t - V_x + U \ast V - V \ast U = 0    \label{dd}
\ee
and
\be
   U_x = {i \over 2} \, (\sigma_3 \, V - V \, \sigma_3) \label{dld} \; .
\ee
Substituting the decomposition 
\be
   V = i \, (V^+ + V^-) \, \sigma_3, \qquad \mbox{with} \qquad
   \sigma_3 \, V^\pm = \pm V^\pm \, \sigma_3
\ee
in (\ref{dld}), we obtain
\be
   V^- = U_x  \; .
\ee
This suggests to impose the condition
\be
   \sigma_3 \, U = - U \, \sigma_3
\ee
on $U$. Inserting this in (\ref{dd}), we obtain $V^+_x = (U\ast U)_x$ and
thus
\be
    V^+ = U \ast U    \label{V+_U}
\ee
up to addition of an arbitrary term which does not depend on $x$ and which 
we disregard in the following. Hence 
\be
    V = i \, (U_x + U \ast U) \, \sigma_3  \; .
\ee
 Furthermore, (\ref{dd}) together with (\ref{V+_U}) leads to
\be
   i \, U_t \, \sigma_3 + U_{xx} - 2 \, U \ast U \ast U = 0  \; .
           \label{mns}
\ee
If we impose the further condition $U^\dag=-U$, then
\be
   U = \left(\begin{array}{cc} 0 & - \bar{\psi} \\ 
                             \psi& 0
			 \end{array} \right)   \qquad
   V = i \left(\begin{array}{cc} - \bar{\psi} \ast \psi & \bar{\psi}_x \\
                                \psi_x & \psi\ast\bar{\psi}
			   \end{array} \right)
\ee
with a complex function $\psi$ and (\ref{mns}) takes the form
\be
   i \, \psi_t + \psi_{xx} + 2 \, \psi \ast \bar{\psi} \ast \psi = 0 \, , 
      \qquad
   i \, \bar{\psi}_t - \bar{\psi}_{xx} - 2 \, \bar{\psi} \ast \psi \ast \bar{\psi} 
   = 0 
\ee
which are the {\em noncommutative nonlinear Schr\"odinger equation} (NNS) 
and its complex conjugate.\footnote{The two equations are complex conjugate 
irrespective of whether $\vartheta$ is real or imaginary. Clearly  
$\overline{f \ast h} = \bar{f} \ast \bar{h}$ if $\vartheta$ 
is real. If $\vartheta$ is imaginary, we have $\overline{f \ast h} = 
\bar{h} \ast \bar{f}$ instead.}
\vskip.2cm

The linear system $\delta \chi = \lambda \, \d \chi$ (where $\delta \chi^{(0)}=0$) 
associated with the NNS reads
\be
   \chi_x = \lambda \, (\chi_t - V \ast \chi) \, ,  \qquad
   (I-\sigma_3) \chi = 2 \, i \, \lambda \, (\chi_x - U \ast \chi) \; .
\ee
Writing $\chi = (\alpha , \beta)^T$, this becomes
\be
   \alpha_x &=& \lambda \, (\alpha_t + i \, \bar{\psi} \ast \psi \ast \alpha
   - i \, \bar{\psi}_x \ast \beta)  \\
   \beta_x &=& \lambda \, (\beta_t - i \, \psi_x \ast \alpha 
   - i \, \psi \ast \bar{\psi} \ast \beta)  \\
   0 &=& \lambda \, (\alpha_x + \bar{\psi} \ast \beta)   \\
   \beta &=& i \, \lambda \, (\beta_x - \psi \ast \alpha)  \; .
\ee
Assuming that $\psi$ is $\ast$-invertible with inverse $\psi_\ast^{-1}$, 
we can use the third equation to eliminate $\beta$ from the first and
the last equation, 
\be 
  \alpha_x &=& \lambda \, (\alpha_t + i \, \bar{\psi} \ast \psi \ast \alpha
   + i \, \bar{\psi}_x \ast \bar{\psi}_\ast^{-1} \ast \alpha_x)  \\
  \alpha_x &=& i \, \lambda \, (\alpha_{xx} - \bar{\psi}_x \ast 
  \bar{\psi}_\ast^{-1} \ast \alpha_x + \bar{\psi} \ast \psi \ast \alpha)
    \; .
\ee
It is not possible now to proceed in perfect analogy with (\ref{alpha_gamma}) 
of the classical case. Instead, we introduce functions $p$ and $q$ such that
\be
   \alpha_t = i \, \lambda \, p \ast \alpha \, , \qquad
   \alpha_x = i \, \lambda \, q \ast \alpha     \label{def_pq}
\ee
assuming that $\alpha$ is $\ast$-invertible. 
In terms of these functions the above equations take the form
\be
   q &=& \bar{\psi} \ast \psi 
   + i \, \lambda \, (q_x - \bar{\psi}_x \ast \bar{\psi}_\ast^{-1} \ast q)
   - \lambda^2 \, q \ast q     \label{q-eq}  \\
   p &=& i \, q_x - 2 \, i \, \bar{\psi}_x \ast \bar{\psi}_\ast^{-1} \ast q
   - \lambda \, q \ast q  \; .  \label{p-eq}
\ee
 From $\alpha_{tx} = \alpha_{xt}$ and (\ref{def_pq}) we find
\be
   q_t - p_x + i \, \lambda \, (q \ast p - p \ast q) = 0 \; .  
                                    \label{cl}
\ee
Using (\ref{ast-comm}) and (\ref{diam}), we find
\be
   q \ast p - p \ast q = \vartheta \, (q_t \diamond p_x - q_x \diamond p_t)
   = \vartheta \, (q \diamond p_x)_t - \vartheta \, (q \diamond p_t)_x 
\ee
where, in the last step, we used the fact that partial derivatives are also 
derivations with respect to the $\diamond$-product.\footnote{Note that, for 
a partial derivative $\partial$, we have 
$\partial \circ m = m \circ \partial_\otimes$ with 
$\partial_\otimes = \partial \otimes 1 + 1 \otimes \partial$ 
which commutes with $P$.}
Now we insert this in (\ref{cl}) and deduce
\be
    w_t = (p + i \, \lambda \, \vartheta \, q \diamond p_t)_x
\ee
where
\be
    w = q + i \, \lambda \, \vartheta \, q \diamond p_x   \; .
           \label{w}
\ee
By expansion into a power series in $\lambda$, this 
yields an infinite set of local conservation laws of the NNS (each of which, 
in turn, can be expanded into a power series in $\vartheta$).
Let us expand $p$ and $q$ into power series in $\lambda$,
\be
    p = \sum_{m=0}^\infty \lambda^m \, p^{(m)} \, , \qquad
    q = \sum_{m=0}^\infty \lambda^m \, q^{(m)} \; .
\ee
Then (\ref{q-eq}) leads to
\be
    q^{(0)} = \bar{\psi} \ast \psi \, , \qquad
    q^{(1)} = i \, \bar{\psi} \ast \psi_x
\ee
and 
\be
  q^{(m)} = i \, ( q^{(m-1)}_x - \bar{\psi}_x \ast \bar{\psi}_\ast^{-1}
            \ast q^{(m-1)} ) - \sum_{k=0}^{m-2} q^{(k)} \ast q^{(m-2-k)} 
\ee
for $m>1$. From (\ref{p-eq}) we get
\be
    p^{(0)} = i \, ( \bar{\psi} \ast \psi_x - \bar{\psi}_x \ast \psi )  
\ee
and 
\be
  p^{(m)} = i \, ( q^{(m)}_x - 2 \, \bar{\psi}_x \ast \bar{\psi}_\ast^{-1}
            \ast q^{(m)} ) - \sum_{k=0}^{m-1} q^{(k)} \ast q^{(m-1-k)} 
\ee
for $m>0$. These formulas allow the recursive calculation of the functions 
$p^{(m)}$ and $q^{(m)}$ in terms of $\psi$. From (\ref{w}) with
$w = \sum_{m \geq 0} w^{(m)}$ we now obtain the following expressions for the 
conserved densities, 
\be
    w^{(0)} &=& \bar{\psi} \ast \psi   \label{w0}   \\
    w^{(1)} &=& i \, \bar{\psi} \ast \psi_x - \vartheta \, (\bar{\psi} \ast \psi)
	            \diamond (\bar{\psi} \ast \psi_{xx} - \bar{\psi}_{xx} \ast \psi)
				                       \label{w1}   \\
    w^{(m)} &=& q^{(m)} + i \, \vartheta \, \sum_{k=0}^{m-1} 
	            q^{(k)} \diamond p^{(m-1-k)}_x   \qquad (m > 1)  \; .
\ee

\section{Single soliton solution of the NNS}
\setcounter{equation}{0}
The one-soliton solution of the classical nonlinear Schr\"odinger equation 
is given by
\be
   \psi = a \, \exp \left( {i \over 2} \, b \, x 
          - i \, ({1 \over 4} \, b^2 - a^2) \, t \right) \,
          \mbox{sech} [ a \, (x-b \, t) ]
\ee
with real constants $a,b$. This expression can be rewritten as follows,
\be
   \psi = a \, F(x) \ast G(x,t) \ast F(x)         \label{nns-1s}
\ee
where
\be
   F(x) = \exp \left( {i \over 2} \, ( {b \over 4} + {a^2 \over b} ) \, x \right) 
          \, , \quad  
   G(x,t) = \exp \left( i \, ( {b \over 4} - {a^2 \over b} ) (x - b \, t) \right) 
            \, \mbox{sech} [ a \, (x-b \, t) ]  \; .
\ee
Clearly, this reduces to the classical solution if $\vartheta = 0$. It is therefore 
sufficient to show that (\ref{nns-1s}) does not depend on $\vartheta$. 
We make use of the identity
\be
  {\pa \over \pa \vartheta} ( f \ast h ) = {\pa f \over \pa \vartheta} \ast h
  + f \ast {\pa h \over \pa \vartheta} + {1 \over 2} ( f_t \ast h_x - f_x \ast h_t )
      \label{vartheta_id}
\ee
for functions $f$ and $h$. Applying this to functions $f(x),g(x,t),h(x)$ which do 
not depend on $\vartheta$, we find 
\be
   {\pa \over \pa \vartheta} ( f \ast g \ast h ) =
 {1 \over 2} ( f \ast g_t \ast h_x - f_x \ast g_t \ast h ) 
\ee
which vanishes if $f_x = c \, f$, $h_x = c \, h$ with a constant $c$ (for arbitrary $g$). 
Since (\ref{nns-1s}) has this special form, it is indeed independent of $\vartheta$.
\vskip.2cm

Our next observation is that a $\ast$-product of two functions is classical if the 
functions depend both on the same argument linear in $x$ and $t$ and if they do not 
depend on $\vartheta$. Hence $f(x-b \, t) \ast h(x-b \, t) = f(x-b \, t) \, h(x-b \, t)$. 
Indeed, by application of (\ref{vartheta_id}) we have
\be
    {\pa \over \pa \vartheta} [ f(x-b \, t) \ast h(x-b \, t ) ]
  = {1 \over 2} ( -b \, f' \ast h' + b \, f' \ast h' ) = 0 
\ee
where $f'$ denotes the derivative of $f$. As a consequence,
\be
      \psi \ast \bar{\psi} \ast \psi 
  &=& a^3 \, F(x) \ast G(x,t) \ast \overline{G(x,t)} \ast G(x,t) \ast F(x)
                                         \nonumber \\
  &=& a^3 \, F(x) \ast 
        \left[ \exp \left( i \, ( {b \over 4} - {a^2 \over b} ) (x - b \, t) \right) 
        \, \mbox{sech}^3 [ a \, (x-b \, t) ] \right] \ast F(x)
                                         \nonumber \\
  &=& |\psi|^2 \, \psi
\ee
using again our previous argument in the last step. 
Therefore the NNS reduces to the classical equation in case of the single soliton solution. 
We have thus shown that the single soliton solution of the classical nonlinear Schr\"odinger
equation remains a solution of the NNS. A similar result should not be expected for 
multi-soliton solutions.

\section{Some perturbative properties of the NNS}
\setcounter{equation}{0}
Expansion of (\ref{ast}) leads to
\be
   f \ast h = f \, h + {\vartheta \over 2} \, (f_t \, h_x - f_x \, h_t)
     + {\vartheta^2 \over 8} \, (f_{tt} \, h_{xx} - 2 \, f_{tx} \, h_{tx}
     + f_{xx} \, h_{tt}) + {\cal O}(\vartheta^3)  \; .
\ee
Since $f$ and $h$ in general depend on $\vartheta$, corresponding power series 
expansions have to be inserted in the last expression and terms rearranged in
ascending powers of $\vartheta$. 
Using this formula, we obtain the following expression for the nonlinear term in 
the NNS,
\be
     \psi \ast \bar{\psi} \ast \psi 
 &=& |\psi|^2 \, \psi + (\vartheta^2/ 4) \, \left( 
   [ \psi_{tt} \psi_{xx} - (\psi_{tx})^2 ] \, \bar{\psi} 
 + 2 \, [ \psi_t \psi_{xx} - \psi_x \psi_{tx} ] \, \bar{\psi}_t 
     \right.             \nonumber \\
 & & 
 + 2 \, [ \psi_{tt} \psi_x - \psi_{tx} \psi_t ] \, \bar{\psi}_x 
 + [ \psi \psi_{xx} - (\psi_x)^2 ] \, \bar{\psi}_{tt} 
 + [ \psi \psi_{tt} - (\psi_t)^2 ] \, \bar{\psi}_{xx}
                         \nonumber \\
 & & \left. 
 + 2 \, [ \psi_t \psi_x - \psi \psi_{tx} ] \, \bar{\psi}_{tx} \right)
 + {\cal O}(\vartheta^3)   \; .
\ee
In particular, the contribution of first order in $\vartheta$ from the 
expansion of the $\ast$-product vanishes identically. By expansion 
of $\psi$, i.e.,
\be
    \psi = \psi_0 + \vartheta \, \psi_1 + {1 \over 2} \, \vartheta^2 \, \psi_2
           + {\cal O}(\vartheta^3) \, ,
\ee
terms linear in $\vartheta$ arise, however. Now the first correction
to the nonlinear Schr\"odinger equation obtained from the NNS is given by
\be
    i \, \psi_{1,t} + \psi_{1,xx} + 4 \, |\psi_0|^2 \, \psi_1 
    \pm 2 \, \psi_0^2 \, \bar{\psi}_1 = 0  
\ee
where the choice of sign depends on whether $\vartheta$ is chosen 
real or imaginary. This equation is linear and homogeneous in $\psi_1$
and thus admits the solution $\psi_1 =0$. As a consequence, every solution 
of the classical nonlinear Schr\"odinger equation is a solution of the NNS 
to first order in $\vartheta$. 
The second correction (quadratic in $\vartheta$) to the nonlinear Schr\"odinger 
equation is an inhomogenous linear equation for $\psi_2$, 
\be
    i \, \psi_{2,t} + \psi_{2,xx} + 2 \, |\psi_0|^2 \, \psi_2 
    + \psi_0^2 \, \bar{\psi}_2 = - 2 \, \psi_1^2 \, \bar{\psi}_0 
    \mp 4 \, \psi_0 \, |\psi_1|^2 - \Gamma/2    
                \label{psi2-eq}
\ee
with
\be
     \Gamma 
 &=& -4 \, \psi_0^4 \, \bar{\psi}_0 \, \bar{\psi}_{0,x}^2 
     + 3\, \psi_{0,xx}^2 \, \bar{\psi}_{0,xx} 
     - 2 \, \psi_{0,x} \, \bar{\psi}_{0,xx} \, \psi_{0,xxx} 
      + \bar{\psi}_0 \, \psi_{0,xxx}^2 
		\nonumber \\
 & & - 4 \, \psi_0^2 \, [ \bar{\psi}_0^3 \, \psi_{0,x}^2 + 
        3 \, \psi_{0,x} \, \bar{\psi}_{0,x} \, \bar{\psi}_{0,xx}  
     + \bar{\psi}_0 \, ( 2\,\psi_{0,xx} \, \bar{\psi}_{0,xx}   
     + \psi_{0,x} \, \bar{\psi}_{0,xxx}  ) ]        
	     \nonumber \\
 & & - 4 \, \psi^3_0 \, \bar{\psi}_{0,x} \,
      ( 2 \, \bar{\psi}_0^2 \, \psi_{0,x} + \bar{\psi}_{0,xxx} )   
     + \psi_{0,xx} \, ( 2\,\bar{\psi}_{0,x} \, \psi_{0,xxx}  
     + 2 \, \psi_{0,x} \, \bar{\psi}_{0,xxx}  
     - \bar{\psi}_0 \,\psi_{0,xxxx} )   
                                         \nonumber \\
 & & - 2\,\psi_{0,x} \, \bar{\psi}_{0,x} \,\psi_{0,xxxx} 
     + \psi_{0,x}^2 \, \bar{\psi}_{0,xxxx} 
     - \psi_0 \, [ 16 \, \bar{\psi}_0 \, \psi_{0,x} \, \bar{\psi}_{0,x} \, \psi_{0,xx} 
     + 4 \, \bar{\psi}_0^2 \, \psi_{0,xx}^2         
                 \nonumber \\
 & & + 4 \, \psi_{0,x}^2 \, ( 3 \, \bar{\psi}_{0,x}^2 + \bar{\psi}_0 \, \bar{\psi}_{0,xx} ) 
     + 2 \, \psi_{0,xxx} \, \bar{\psi}_{0,xxx} +  \bar{\psi}_{0,xx} \, \psi_{0,xxxx}  
     + \psi_{0,xx} \, \bar{\psi}_{0,xxxx} ]  
\ee
where we used that $\psi_0$ satisfies the classical nonlinear 
Schr\"odinger equation.
\vskip.2cm

Expanding the expressions for the conserved densities $w^{(0)}$ and 
$w^{(1)}$ (see (\ref{w0}) and (\ref{w1})) in powers of $\vartheta$, 
we obtain
\be
   w^{(0)} &=& \bar{\psi}_0 \, \psi_0 + \vartheta \, [ \bar{\psi}_0 \psi_1 
               + \bar{\psi}_1 \psi_0 + {1 \over 2} \,
               (\bar{\psi}_{0,t} \, \psi_{0,x} - \bar{\psi}_{0,x} \, \psi_{0,t}) ] 
               + {\cal O}(\vartheta^2)       \nonumber       \\   
		   &=& \bar{\psi}_0 \, \psi_0 + \vartheta \, [ \bar{\psi}_0 \psi_1 
		       + \bar{\psi}_1 \psi_0 - {i \over 2} \,
		       ( \bar{\psi}_{0,x} \, \psi_{0,x} + |\psi_0|^4 )_x ]
               + {\cal O}(\vartheta^2) \\
   w^{(1)} &=& i \, \bar{\psi}_0 \, \psi_{0,x} + \vartheta \, 
               [ i \, (\bar{\psi}_0 \, \psi_{1,x} + \bar{\psi}_1 \, \psi_{0,x})
			          \nonumber \\
		   & & + {i \over 2} \, (\bar{\psi}_{0,t} \, \psi_{0,xx}
               - \bar{\psi}_{0,x} \, \psi_{0,xt}) - |\psi_0|^2 \, 
               (\bar{\psi}_0 \, \psi_{0,xx} - \bar{\psi}_{0,xx} \, \psi_0) ]
               + {\cal O}(\vartheta^2)     \nonumber \\
		   &=& i \, \bar{\psi}_0 \, \psi_{0,x} + \vartheta \, 
		       [ i \, (\bar{\psi}_0 \, \psi_{1,x} + \bar{\psi}_1 \, \psi_{0,x})
			   + {i \over 2} \, (\bar{\psi}_{0,x} \, \psi_{0,xx}
		       + \bar{\psi}_0 \, \bar{\psi}_{0,x} \, \psi_0^2 )_x ]
               + {\cal O}(\vartheta^2)    \; .
\ee
In the last step we used again that $\psi_0$ solves the classical nonlinear 
Schr\"odinger equation. Hence, those parts of the first order corrections to 
the conserved densities which do not depend on $\psi_1$ are total 
$x$-derivatives and thus do not contribute to the conserved charges. 
A corresponding evaluation of the second order corrections is much more 
complicated since one has to solve the inhomogeneous equation (\ref{psi2-eq}).

\section{Conclusions}
\setcounter{equation}{0}
By deformation quantization applied to a bicomplex associated with the 
nonlinear Schr\"odinger equation, we obtained a deformed equation (NNS) which 
lives on a noncommutative space-time and which shares with its classical 
version the property of having an infinite set of conserved quantities, a 
characteristic feature of soliton equations and (infinite-dimensional) 
integrable models. Surprisingly, the one-soliton solution of the classical 
equation remains a solution of the NNS. The fate of the classical multi-soliton 
structure under the deformation has still to be explored.
\vskip.2cm

We expect that in a similar way interesting noncommutative versions 
of many other integrable models can be constructed (see also \cite{DMH00b}).


\begin{thebibliography}{**}
\bibitem{ncg_str} Connes A., Douglas M. and Schwarz A., Noncommutative
 geometry and matrix theory: compactification on tori, {\em JHEP}, 1998, V.02,
 003 \\
 Douglas M. R. and Hull C., D-branes and  the noncommutative torus,
 {\em JHEP}, 1998, V.02, 008 \\
 Ardalan F., Arfaei H. and Sheikh-Jabbari M. M., Noncommutative geometry from
 strings and branes, {\em JHEP}, 1999, V.02, 016 \\
 Cheung Y.-K. E. and Krogh M., Noncommutative geometry from D0-branes in a
 background B-field, {\em Nucl. Phys. B}, 1998, V.528, 185-196 \\
 Seiberg N. and Witten E., String theory and noncommutative geometry, 
 {\em JHEP}, 1999, V.09, 032.
\bibitem{dq} Bayen F., Flato M., Fronsdal C., Lichnerowicz A. and Sternheimer D.,
 Deformation theory and quantization I, II {\em Ann. Phys.}, 1978, V.111, 
 61--151.
\bibitem{dq_nc}
 Andreev O. and Dorn H., On open string sigma-model and noncommutative
 gauge fields, hep-th/9912070 \\
 Hashimoto K. and Hirayama T., Branes and BPS configurations of
 non-commutative/commutative gauge theories, hep-th/0002090 \\
 Asakawa T. and Kishimoto I., Noncommutative gauge theories
 from deformation quantization, hep-th/0002138 \\
 Gopakumar R., Minwalla S. and Strominger A.,
 Noncommutative solitons, hep-th/0003160 \\
 Moriyama S., Noncommutative monopole from nonlinear monopole,
 hep-th/0003231 \\
 Goto S. and Hata H., Noncommutative monopole at the second order in $\theta$,
 hep-th/0005101 \\ 
 Gross D. J. and Nekrasov N. A., Monopoles and strings in noncommutative gauge theory,
 hep-th/0005204 \\
 Ydri B., Noncommutative geometry as a regulator, hep-th/0003232 \\
 Chamseddine A. H., Complexified gravity in noncommutative spaces, 
 hep-th/0005222  \\ 
 Girotti H. O., Gomes M., Rivelles V. O. and da Silva A. J., A consistent noncommutative 
 field theory: the Wess-Zumino model, hep-th/0005272 \\
 Periwal V., Nonperturbative effects in deformation quantization, hep-th/0006001.
\bibitem{def_im} 
 Strachan I. A. B., The Moyal bracket and the dispersionless limit of the KP hierarchy
 {\em J.Phys. A}, 1995, V.28, 1967--1976; The dispersive self-dual Einstein equations 
 and the Toda lattice {\em J.Phys. A}, 1996, V.29, 6117--6124 \\
 Castro C., Large Nonlinear $W_{\infty}$ algebras from nonlinear integrable deformations
 of self dual gravity {\em Phys. Lett. B}, 1995, V.353, 201--208 \\
 Takasaki K., Dressing operator approach to Moyal algebraic deformation of 
 selfdual gravity {\em J. Geom. Phys.}, 1994, V.14, 111--120 \\
 Kemmoku R., Difference operator approach to the Moyal quantization and its 
 application to integrable systems, {\em J. Phys. Soc. Japan}, 1997, V.66, 51--59 \\
 Garc{\'i}a-Compe{\'a}n H., Pleba{\'n}ski J. F. and Przanowski M., 
 Geometry associated with self-dual Yang-Mills and the chiral model 
 approaches to self-dual gravity {\em Acta Phys. Polon. B}, 1998, V.29, 549--571 \\
 Castro C. and Plebanski J., The generalized Moyal Nahm and continuous Moyal Toda 
 equations {\em J. Math. Phys.}, 1999, V.40, 3738--3760.
\bibitem{Stra97} Strachan I. A. B., A geometry of multidimensional integrable systems 
 {\em J. Geom. Phys.}, 1997, V.21, 255--278.
\bibitem{nc_im} 
 Takasaki K., Anti-self-dual Yang-Mills equations on noncommutative 
 spacetime, hep-th/0005194.
\bibitem{DMH00a} Dimakis A. and M\"uller-Hoissen F., 
 Bi-differential calculi and integrable models  
 {\em J. Phys. A: Math. Gen.}, 2000, V.33, 957--974;
 Bicomplexes and integrable models, nlin.SI/0006029.
\bibitem{Fadd+Takh87} Faddeev L. D. and Takhtajan L. A.,
 {\em Hamiltonian Methods in the Theory of Solitons}, Springer, Berlin, 1987.
\bibitem{DMH00b} Dimakis A. and M\"uller-Hoissen F., 
 Bicomplexes, integrable models, and noncommutative geometry, hep-th/0006005. 
\end{thebibliography}
\end{document}